\documentclass[10pt,aps,prd,reprint,nofootinbib,floats,floatfix,amsfonts,amssymb,amsmath,preprintnumbers,notitlepage,superscriptaddress,twocolumn]{revtex4-1}

\usepackage{graphicx,xcolor} % Figure support
\usepackage{hyphenat} % Allow line break on hyphenated words
\usepackage{amsmath, amssymb, amsfonts,amsbsy,amsthm} %Math and such
\usepackage{float}
\usepackage{subcaption}
\usepackage{graphicx}
\usepackage{ulem}
\usepackage{units}

\newcommand{\orcid}[1]{\href{https://orcid.org/#1}{\includegraphics[scale=0.15]{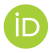}}}

\begin{document}

\title{Formalism for power spectral density estimation for non-identical and correlated noise using the null channel in Einstein Telescope}

\author{Kamiel Janssens\orcid{0000-0001-8760-4429}}
\affiliation{Universiteit Antwerpen, Prinsstraat 13, 2000 Antwerpen, Belgium}
\affiliation{Universit\'e C$\hat{o}$te d’Azur, Observatoire C$\hat{o}$te d’Azur, CNRS, Artemis, F-06304 Nice, France}

\author{Guillaume Boileau \orcid{0000-0002-3576-69689}}
\affiliation{Universiteit Antwerpen, Prinsstraat 13, 2000 Antwerpen, Belgium}

\author{Marie-Anne~Bizouard\orcid{0000-0002-4618-1674}}
\affiliation{Universit\'e C$\hat{o}$te d’Azur, Observatoire C$\hat{o}$te d’Azur, CNRS, Artemis, F-06304 Nice, France}

\author{Nelson Christensen\orcid{0000-0002-6870-4202}}
\affiliation{Universit\'e C$\hat{o}$te d’Azur, Observatoire C$\hat{o}$te d’Azur, CNRS, Artemis, F-06304 Nice, France}

\author{Tania Regimbau\orcid{0000-0002-0631-1198}}
\affiliation{Laboratoire d’Annecy de Physique des Particules, CNRS, 9 Chemin de Bellevue, 74941 Annecy, France
}

\author{Nick van Remortel\orcid{0000-0003-4180-8199}}
\affiliation{Universiteit Antwerpen, Prinsstraat 13, 2000 Antwerpen, Belgium}

\date{\today}

\begin{abstract}

Several proposed gravitational wave interferometers have a triangular configuration, such as the Einstein Telescope and the Laser Interferometer Space Antenna. For such a configuration one can construct a unique null channel insensitive to gravitational waves from all directions. We expand on earlier work and describe how to use the null channel formalism to estimate the power spectral density for the Einstein Telescope interferometers with non-identical as well as correlated noise sources. The formalism is illustrated with two examples in the context of the Einstein Telescope, with increasing degrees of complexity and realism. By using known mixtures of noises we show the formalism is mathematically correct and internally consistent. Finally we highlight future research needed to use this formalism as an ingredient for a Bayesian estimation framework.
\end{abstract}

\maketitle

%%%% Main text %%%%

%%%%%%%%%%%
\section{Introduction}
\label{sec:Introduction}
%%%%%%%%%%%

The Einstein Telescope (ET) \cite{Punturo:2010zz} and Cosmic Explorer (CE) \cite{Reitze2019Cosmic,Evans:2021gyd} are respectively the European and U.S. proposals for third generation Earth-based interferometric gravitational-wave (GW) detectors. They are planned to outperform the current generation of gravitational wave (GW) interferometers, LIGO \cite{2015}, Virgo \cite{VIRGO:2014yos} and KAGRA \cite{PhysRevD.88.043007}, by an order of magnitude in strain sensitivity and to be sensitive in a wider frequency band. The current proposal for ET consists of an equilateral triangle built-up with six interferometers (consisting of three for low frequency, three for high frequency), with an opening angle of $\pi/3$ and arm lengths of 10 km. In the rest of this paper, we ignore the details of the xylophone configuration and treat ET as three interferometers~\cite{Hild:2010id}. The CE is planned to be an L-shaped interferometer with an arm length of 40 km.

The future space-based GW detector Laser Interferometer Space Antenna (LISA) \cite{2017arXiv170200786A}, has also an equilateral triangular configuration, however the length of the arms is \unit[$2.5 \times 10^6$]{km}. Whereas ET and CE will be sensitive to GWs with frequencies of a couple of Hz to a couple of kHz, LISA is sensitive to GWs with frequencies between \unit[0.01]{mHz} and \unit[1]{Hz}. Other proposed triangular GW detectors are DECIGO~\cite{Sato_2017} and TianQin~\cite{TianQin:2015yph}.

Given the expected sensitivities for ET, CE and LISA, it is predicted they will observe a large number of overlapping signals \cite{PhysRevD.86.122001,PhysRevD.79.062002,2017arXiv170200786A}. The constant presence of numerous signals will make the estimation of the noise power spectral density (PSD) of the these interferometers challenging, while unbiased noise PSD estimation is a key ingredient for all GW detection pipelines.

Another issue that one might face when analyzing the data of ET and LISA is the presence of non-negligible amounts of correlated noise between the different interferometers, due to (almost) co-located input/output test masses of the different interferometers. 
Several studies focusing on the ET have pointed out to the possibility of correlated seismic and Newtonian noise \cite{PhysRevD.106.042008} as well as magnetic noise \cite{PhysRevD.104.122006,https://doi.org/10.48550/arxiv.2209.00284}. For LISA correlated effects from micro-thrusters, the magnetic field and temperature variations on the test-mass system can be expected~\cite{2022arXiv220403867B}.
Correlated noise is particularly problematic for unmodeled sources of GWs which depend on cross correlation methods, i.e. the search for a stochastic gravitational wave background (SGWB) and searches for poorly known transient signals -- called $bursts$ -- such as core collapse supernovae. 

A method often considered for PSD estimation is using the null channel\footnote{The null channel is also know as the null stream or Sagnac channel.}. For a network consisting of $N$ detectors one can construct $N - 2$ sky location dependent null streams \cite{PhysRevD.40.3884,Wen_2005,Wen_2008,PhysRevD.81.082001,Sutton:2009gi}. However, in the case of a triangular configuration of three interferometers, there is one unique null channel which is insensitive to GWs from every direction \cite{PhysRevD.86.122001,Tinto2005,PhysRevD.100.104055,2022PhRvD.105h4002W}. 
Due to imperfect knowledge on the sky location of the source it is difficult to use the sky location dependent null channel for noise PSD estimation. For future detectors such as ET, CE and LISA, it is even practically impossible due to the high probability of overlapping signals. As soon as there are overlapping signals present in the detector the sky location dependent null channel is unable to remove the multiple signals simultaneously due to their different sky locations. In the remainder of the paper we refer to the sky location independent null channel as 'null channel'.

Previous work has discussed the null channel for PSD estimation with the ET \cite{PhysRevD.86.122001} in the case of identical and uncorrelated noise for the three different interferometers. A more recent study investigates the effectiveness of the null channel in the presence of noise transients, also known as glitches \cite{PhysRevD.105.122007}. Also the authors of \cite{2022PhRvD.105h4002W} describe the benefit of using a null and signal space for parameter estimation in the context of the ET. The consideration of using signal and null space is also used in the context of LISA \cite{PhysRevD.82.022002, 2014PhDT.......286A,2021PhRvD.103j3529B,2021MNRAS.508..803B,2022PhRvD.105b3510B,PhysRevD.103.042006}. Some of these works investigate the possibility of correlated noise \cite{PhysRevD.82.022002, 2014PhDT.......286A,2021PhRvD.103j3529B,2021MNRAS.508..803B,2022PhRvD.105b3510B} and other work of non-identical noise \cite{PhysRevD.103.042006}.

In this paper we expand on these previous studies by discussing how to perform PSD estimation in the case of correlated and non-identical noise. Whereas earlier work has investigated non identical \cite{PhysRevD.105.122007,PhysRevD.82.022002, 2014PhDT.......286A,2021PhRvD.103j3529B,2021MNRAS.508..803B,2022PhRvD.105b3510B} and correlated noise \cite{PhysRevD.103.042006} separately, we present one formalism to address both non-identical and correlated noise simultaneously.
While such a formalism is needed to achieve unbiased PSD estimations, it also provides further information on the correlated noise sources, which is interesting in its own right, particularly for SGWB and burst searches. 

After defining the null channel in Sec.\ref{sec:nullchannel} we derive the formula of the power and cross-power spectral densities for the Einstein Telescope interferometers in Sec.\ref{sec:nullchannel_complex}. The use of the formalism is illustrated with two noise examples in Sec.\ref{sec:demonstration} while we draft in Sec.\ref{sec:Outlook} the corresponding Bayesian statistical estimation framework.

%%%%%%%%%%%
\section{The null channel}
\label{sec:nullchannel}
%%%%%%%%%%%

The null channel is the sum of the strain output of three interferometers in an equilateral triangle, which we will call $X$, $Y$ and $Z$. Each detector measures a strain time series $s^I(t)$, which consists both of noise $n^I(t)$ as well as a GW component $h^I(t)$,
\begin{equation}
\label{eq:xa}
    s^{ I}(t) \equiv n^{ I}(t) +h^{ I}(t),
\end{equation}
where $I$ runs over $X$, $Y$ and $Z$. The null channel is given by the sum of the output of the three interferometers \cite{PhysRevD.86.122001}:

\begin{equation}
\label{eq:nullchannel}
    \begin{aligned}
        s_{null}(t) &\equiv   \sum_{I= X}^{Z} s^I(t)\\
        & =  \sum_{I= X}^{Z} n^{ I}(t) + \sum_{I= X}^{Z} h^{ I}(t)\\
        & = \sum_{I= X}^{Z} n^{ I}(t)
    \end{aligned}
\end{equation}
The derivation of $\sum_{I= X}^{Z} h^{ I}(t) = 0 $ for three  interferometers in an equilateral triangle configuration is discussed in \cite{PhysRevD.86.122001,2022PhRvD.105h4002W}. The equality holds for any polarisation of GW since the sum of the detector response functions of the interferometers is zero. However the calculation assumes the arms of the different interferometers are exactely co-located. For the ET, deviations can be expected since the terminal and central stations of two different ET interferometers are separated by about \unit[300--500]{m} \cite{ETdesignRep}. In this paper we do not consider any of such deviations and assume the interferometers to be exactly co-located, leading to a perfect null channel.
Furthermore the null channel deteriorates for higher frequencies in an interferometric gravitational-wave detector. This is due to finite arm length
effects. There are further imperfections when the arm lengths
are not exactly equal, as will be the case for LISA \cite{PhysRevD.82.022002}

We introduce the PSD of the strain of interferometer $I$, $S^I_{s}$,

\begin{equation}
\label{eq:xa_PSD}
    \begin{aligned}
    \langle s^{I}(f) s^{I,*}(f^{\prime})\rangle &= \frac{1}{2}\delta(f-f^{\prime}) S^I_{s}(f) \\
    & = \frac{1}{2}\delta(f-f^{\prime}) \left[ S^I_{n}(f) + S^I_{h}(f) \right],
    \end{aligned}
\end{equation}
where $S^I_{n}(f)$ and $S^I_{h}(f)$ are respectively the noise and GW PSDs for interferometer $I$. The cross spectral density (CSD) of the strain of interferometers $I$ and $J$, $S^{IJ}_{s}(f)$, is given by

\begin{equation}
\label{eq:xa_CSD}
    \begin{aligned}
    \langle s^{I}(f) s^{J,*}(f^{\prime}) \rangle &= \frac{1}{2}\delta(f-f^{\prime}) S^{IJ}_{s}(f) \\
    & = \frac{1}{2}\delta(f-f^{\prime}) \left[ S^{IJ}_{n}(f) + S^ {IJ}_{h}(f) \right],
    \end{aligned}
\end{equation}
where $S^{IJ}_{n}(f)$ and $S^{IJ}_{h}(f)$ are respectively the noise and GW CSDs for interferometers $I$ and $J$.

The quantities $S^I_{h}(f)$ and $S^{IJ}_{h}(f)$ depend on the response of the interferometer(s) $I$ and respectively $I$ and $J$, whereas one typically is interested in $S_{h}(f)$ of the source regardless of the observing interferometer. In this paper, we focus on an isotropic SGWB with equal levels of tensor cross- and plus- polarization. However, the formalism we formulate in Sec.~\ref{sec:nullchannel_complex} can be used to estimate noise PSDs in the presence of any other GW signal. In Appendix \ref{sec:Appendix1-T} the following equalities are derived for an isotropic SGWB with equal levels of tensor cross- and plus- polarization,

\begin{equation}
\label{eq:sh}
    \begin{aligned}
    S^I_{h}(f) &= \frac{3}{10} S_{h}(f) \\
    S^{IJ}_{h}(f) & = - \frac{3}{20} S_{h}(f).
    \end{aligned}
\end{equation}
In Appendix \ref{sec:Appendix2-SV} we also derive the equivalent terms in case of the presence of an isotropic SGWB with scalar or vector polarizations, which are predicted by several extensions of general relativity~\cite{Callister:2017ocg}.

One can already use the null channel for getting information on the detector noise, but even more information can be extracted using a set of three channels $A$, $E$ and $T$. These three channels are often used in the context of LISA and are defined as the following linear combinations of the three $X$, $Y$ and $Z$ interferometers \cite{Tinto2005, PhysRevD.100.104055}: 
\begin{equation}
\label{eq:AETchannels}
    \begin{aligned}
    A &= \frac{1}{\sqrt{2}} (Z-X)    \\
    E &= \frac{1}{\sqrt{6}} (X-2Y+Z) \\
    T &= \frac{1}{\sqrt{3}} (X+Y+Z).
    \end{aligned}
\end{equation}

The $T$ channel is a normalized version of the null channel as first defined in Eq.~\ref{eq:nullchannel} and is insensitive to GWs, whereas the $A$ and $E$ channels contain the GW signal \cite{PhysRevD.100.104055}. Under the assumption of identical noise sources in $X$, $Y$ and $Z$, the $A$, $E$ and $T$ channels are orthogonal with respect to one another, implying that their noise PSDs are uncorrelated.  

Note that the use of the $A$, $E$ and $T$ channels as used by the LISA community is equivalent to the description of the null and signal space in the context of the ET \cite{2022PhRvD.105h4002W}. The signal space considered in \cite{2022PhRvD.105h4002W} consists of two channels which correspond to $-A$ and $-E$ in this work and the null space corresponds to $T$.

%%%%%%%%%%%
\section{Null channel in a complex environment}
\label{sec:nullchannel_complex}
%%%%%%%%%%%

In this section, we introduce a formalism to expand on the null channel by using the $A$, $E$ and $T$ channels. First, we discuss the presence of identical correlated noise, which already has been studied in the context of LISA~\citep{PhysRevD.66.122002}. Afterwards, we relax this assumption and describe the most general case describing non-identical and correlated noise sources.

\subsection{Formalism in the presence of identical correlated noise}

In the presence of correlated noise, identical in all interferometers -- $S_n^{XY}(f) = S_n^{XZ}(f) = S_n^{YZ}(f) \equiv S_n^{IJ}(f)$ -- as well as assuming identical noise PSDs -- $S_n^{X}(f) = S_n^{Y}(f) = S_n^{Z}(f) \equiv S_n^{I}(f)$ --  the following equalities hold for the $A$, $E$ and $T$ channels
\begin{widetext}
\begin{equation}
\begin{aligned}
\label{eq:identicalSummary}
    \langle T(f) T^*(f^{\prime})\rangle &=   \frac{1}{2}\delta(f-f^{\prime}) \left[ S_n^{I}(f) +2S_n^{IJ}(f) \right]\\ 
    \langle A(f) A^*(f^{\prime})\rangle &=   \frac{1}{2}\delta(f-f^{\prime}) \left[ S_n^{I}(f) -S_n^{IJ}(f) +  \mathcal{O}\left(S_h(f)\right) \right] \\
    \langle E(f) E^*(f^{\prime})\rangle &=   \frac{1}{2}\delta(f-f^{\prime}) \left[ S_n^{I}(f) -S_n^{IJ}(f) + \mathcal{O}\left(S_h(f)\right) \right]  \\
    \langle T(f) A^*(f^{\prime})\rangle = \langle A(f) T^*(f^{\prime})\rangle &=  0 \\    
    \langle T(f) E^*(f^{\prime})\rangle = \langle E(f) T^*(f^{\prime})\rangle &=  0 \\
    \langle E(f) A^*(f^{\prime})\rangle = \langle A(f) E^*(f^{\prime})\rangle &=   \mathcal{O}\left(S_h(f)\right). \\     
\end{aligned}
\end{equation}
\end{widetext}
The normalized null channel, i.e. the $T$ channel, can be used for noise PSD estimation since it is independent of the GW signal. 

We note there is no correlation between any of the $A$, $E$ and $T$ channels, since the channels were constructed to form independent signal and null space \cite{Tinto2005, PhysRevD.100.104055,2022PhRvD.105h4002W}.
The exact value of the GW term $\mathcal{O}\left(S_h(f)\right)$ for an isotropic SGWB is derived in Appendix \ref{sec:Appendix1-T} and Appendix \ref{sec:Appendix2-SV} for respectively Tensor and Scalar-Vector polarized GWs.

\subsection{Formalism in the presence of non-identical and correlated noise} 

In the more general scenario where the correlated noise as well as the noise PSDs are unique for each interferometer, Eq. \ref{eq:identicalSummary} changes to, 
\begin{widetext}
\begin{equation}
\begin{aligned}
\label{eq:differentSummary}
    \langle T(f) T^*(f^{\prime})\rangle =  \frac{1}{2}\delta(f-f^{\prime})  &\left[\frac{1}{3} (S_n^X(f) +  S_n^Y(f) +  S_n^Z(f)) + \frac{2}{3} (S_n^{XY}(f) +  S_n^{XZ}(f) +  S_n^{YZ}(f)) \right] \\
    \langle A(f) A^*(f^{\prime})\rangle =  \frac{1}{2}\delta(f-f^{\prime})  &\left[ \frac{1}{2} ( S_n^X(f) + S_n^Z(f))  - S_n^{XZ}(f) +  \mathcal{O}\left(S_h(f)\right) \right] \\
    \langle E(f) E^*(f^{\prime})\rangle =   \frac{1}{2}\delta(f-f^{\prime})  &\left[\frac{1}{6} ( S_n^X(f) + 4 S_n^Y(f) + S_n^Z(f)) + \frac{1}{3} S_n^{XZ}(f) - \frac{2}{3}(S_n^{XY}(f)+S_n^{YZ}(f)) \right. \\
    & \left. \mathcal{O}\left(S_h(f)\right)   \right]  \\
    \langle T(f) A^*(f^{\prime})\rangle = \langle A(f) T^*(f^{\prime})\rangle =  \frac{1}{2}\delta(f-f^{\prime}) &\left[\frac{1}{\sqrt{6}}( S_n^Z(f) - S_n^X(f) + S_n^{YZ}(f) - S_n^{XY}(f) ) \right]  \\
    \langle T(f) E^*(f^{\prime})\rangle = \langle E(f) T^*(f^{\prime})\rangle =  \frac{1}{2}\delta(f-f^{\prime}) & \left[ \frac{1}{3\sqrt{2}} (S_n^X(f) - 2 S_n^Y(f) + S_n^Z(f) - S_n^{XY}(f) + 2 S_n^{XZ}(f) - S_n^{YZ}(f) ) \right]  \\     
    \langle E(f) A^*(f^{\prime})\rangle = \langle A(f) E^*(f^{\prime})\rangle = \frac{1}{2}\delta(f-f^{\prime}) & \left[ \frac{1}{2\sqrt{3}} (- S_n^X(f) + S_n^Z(f) - 2 S_n^{YZ}(f) + 2 S_n^{XY}(f) ) + \mathcal{O}\left(S_h(f)\right) \right], \\ 
\end{aligned}
\end{equation}
\end{widetext}
whereas just as in Eq. \ref{eq:identicalSummary} the exact contribution of the GW term $\mathcal{O}\left(S_h(f)\right)$ is derived in Appendix \ref{sec:Appendix1-T} and Appendix \ref{sec:Appendix2-SV} for an isotropic SGWB with different polarisation's. Eq. \ref{eq:differentSummary} presents six independent equations for seven unknowns ($S_n^X(f)$, $S_n^Y(f)$, $S_n^Z(f))$, $S_n^{XY}(f)$, $S_n^{XZ}(f)$, $S_n^{YZ}(f)$ and $S_h(f)$). We want to point out that the frequency behaviour of each of these unknowns is typically parameterized by multiple parameters and therefore the total number of parameters can become very large.

Of these six equations only three are independent of any GW signal and can yield a PSD estimate without the need to rely on the knowledge of the present GW signals. That is, $\langle T(f) T ^*(f^{\prime})\rangle$, $\langle T(f) A ^*(f^{\prime})\rangle$ and $\langle T(f) E ^*(f^{\prime})\rangle$. This shows the advantage of using the $A$, $E$, $T$ channels over using only the $T$-channel, namely the gain of two additional independent equations which are free of any GW signal.

Alternatively on can use another set of three independent equations, free of GWs, given by,
\begin{widetext}
\begin{equation}
\begin{aligned}
\label{eq:differentSummary_TXYZ}
   \langle T(f) X^*(f^{\prime})\rangle = \langle X(f) T^*(f^{\prime})\rangle =  \frac{1}{2}\delta(f-f^{\prime}) & \left[\frac{1}{\sqrt{3}}( S_n^X(f) + S_n^{XY}(f) + S_n^{XZ}(f) ) \right]  \\ 
   \langle T(f) Y^*(f^{\prime})\rangle = \langle Y(f) T^*(f^{\prime})\rangle =  \frac{1}{2}\delta(f-f^{\prime}) &\left[\frac{1}{\sqrt{3}}( S_n^Y(f) + S_n^{YX}(f) + S_n^{YZ}(f) ) \right]  \\ 
    \langle T(f) Z^*(f^{\prime})\rangle = \langle Z(f) T^*(f^{\prime})\rangle =  \frac{1}{2}\delta(f-f^{\prime}) & \left[\frac{1}{\sqrt{3}}( S_n^Z(f) + S_n^{ZX}(f) + S_n^{ZY}(f) ) \right].\\     
\end{aligned}
\end{equation}
\end{widetext}
The use of $\langle T(f) X^*(f^{\prime})\rangle$, $\langle T(f) Y^*(f^{\prime})\rangle$ and $\langle T(f) Z^*(f^{\prime})\rangle$ was proposed in \cite{PhysRevD.105.122007}, mainly in the context of glitch mitigation in compact binary coalescence (CBC) searches and assuming non-correlated noise. In Eq. \ref{eq:differentSummary_TXYZ} we show how one can correctly include correlated noise terms such that an unbiased estimate of the $X$, $Y$ and $Z$ interferometers PSDs can be achieved. 

The equations describing $\langle T(f) X^*(f^{\prime})\rangle$, $\langle T(f) Y^*(f^{\prime})\rangle$ and $\langle T(f) Z^*(f^{\prime})\rangle$ are not independent of the equations describing $\langle T(f) T ^*(f^{\prime})\rangle$, $\langle T(f) A ^*(f^{\prime})\rangle$ and $\langle T(f) E ^*(f^{\prime})\rangle$, since the $A$, $E$ and $T$ channels are linear combinations of the $X$, $Y$ and $Z$ channels. However, the two different sets of equations present a different representation of the same problem and depending on the situation, one or the other might be more or less useful. As an example, Eq. \ref{eq:differentSummary} gives a set of six independent equations that can be used to jointly estimate signal and noise terms. Furthermore the coherence between the $T$ and $A$ and/or $E$ channels can be used as a simple diagnostic tool. In case the coherence is consistent with Gaussian noise this indicates the noise is identical between the $X$, $Y$ and $Z$ interferometers. By correlating the $T$ and $A$ and/or $E$ channels over long timescales, one might also gain additional information on non-identical correlated noise sources which do not directly affect the detectors noise PSD, but might be affecting correlation based searches, such as the search for a SGWB.
On the other hand, Eq. \ref{eq:differentSummary_TXYZ}, might be easier to use if one is just interested in the noise quantities of the $X$, $Y$ and $Z$ interferometers.

In Sec.~\ref{sec:demonstration} we illustrate the formalism using two examples of noise for the ET, whereas in Sec.~\ref{sec:Outlook} we outline the next steps needed to use this formalism in a Bayesian estimation framework.

%%%%%%%%%%%%%
\section{Examples in the context of the Einstein Telescope}
\label{sec:demonstration}
%%%%%%%%%%%%%

In this section, we show how Eq.~\ref{eq:differentSummary} and Eq.~\ref{eq:differentSummary_TXYZ} are used to provide unbiased PSD estimates considering two examples of noise in the Einstein Telescope. Both examples contain a SGWB signal coming from the superposition of CBC events. The first example additionally contains correlated noise sources having a Gaussian spectrum at three different frequencies. Furthermore, to simulate the case of non identical noise spectrum in X, Y and Z, each of the 3 Gaussian noise sources is injected in only two interferometers.
The second example presents a more realistic scenario where the same SGWB is present as well as correlated magnetic and Newtonian noise (NN). The data sets used are described in Sec. \ref{sec:demonstration:datasets}.

The two examples also demonstrate that, although it is insensitive to GWs, the $T$ channel PSD $S_n^{T}$ \footnote{Note that we have chosen to also give a lower index n to the spectral density of the T channel to indicate it only depends on noise terms and is insensitive to GWs.}, as well as the cross-correlated power between the null channel and each interferometer $S_n^{TX}$, $S_n^{TY}$ and $S_n^{TZ}$ yield biased estimates of $S_n^{X}$, $S_n^{Y}$ and $S_n^{Z}$ in the case of non-identical and correlated noise as stated in \cite{PhysRevD.86.122001}. Instead, we demonstrate how the formalism described in this paper can be used to properly take the non-identical and correlated noise into account, such that unbiased estimates can be defined.

We would like to note that whereas $S_n^{T}=S_n^{X}$ in case of identical and non correlated noise in the three interferometers, this is not the case for $S_n^{TX}$, $S_n^{TY}$ and $S_n^{TZ}$ due to the used normalisation. Therefore we use in the remainder of the paper $S_n^{\prime}{}^{TX}=\sqrt{3}S_n^{TX}$ and equivalent for the other channels.

\subsection{Data sets}
\label{sec:demonstration:datasets}

\subsubsection{Model of a SGWB}
\label{sec:datasets:SGWB}

As signal, we use a background formed by the superposition of a population of binary neutron stars distributed isotropically in the sky and up to a redshift of $z=10$. 
For the generation of this SGWB, we rely on Monte Carlo techniques using the MDC\_Generation code \cite{PhysRevD.86.122001} that randomly selects the set of parameters for each individual BNS (coalescence time, position in the sky, merger redshift, component masses and spins, orientation) and produces the corresponding gravitational-wave polarization $h_+(t)$ and $h_\times(t)$. The interferometer response $h^J(t)=h_+(t) F^J_+(t) + h_\times(t) F^J_\times(t)$ is added to the time-domain strain amplitude of the three ET detectors in a frequency band \unit[5]{Hz}-\unit[2048]{Hz}.
For simplicity we assume identical component masses of 1.4 M$_\odot$ and no spin. We also assume that sources are in average separated by a short time interval of $\Delta t$=\unit[1.5]{s}. Although unrealistic this short interval has the advantage to give a continuous and Gaussian background that is detectable within a short observation period. 

To reveal the SGWB, one typically wants to measure its normalized energy density, expressed by \citep{Allen_102001,Romano:2016dpx}
\begin{equation}
    \label{eq:omgeaGWB}
    \Omega_{\rm GW}(f) = \frac{1}{\rho_{\rm c}}\frac{\text{d}\rho_{\rm GW}}{\text{d}\ln f}~,
\end{equation}
where $\text{d}\rho_{\rm GW}$, is the energy density contained in a logarithmic frequency interval, $\text{d}\ln f$.  Here, $\rho_{\rm c} = 3H_0^2c^2/(8\pi G)$ is the critical energy density for a flat Universe.
In the absence of correlated noise one can construct a cross-correlation statistic which is an unbiased estimator of $\Omega_{\rm GW}(f)$ as follows,
\begin{equation}
    \label{eq:cross-correlationstatistic}
\hat{C}_{IJ}(f) = \frac{2}{\Delta T} \frac{{\rm{Re}}[\Tilde{s}^*_I(f)\Tilde{s}_J(f)]}{\gamma_{IJ}(f)S_0(f)}~,
\end{equation}
for interferometers $I$ and $J$, where
$\Tilde{s}_I(f)$ is the Fourier transform of the time domain strain data $s_I(t)$ measured by interferometer $I$, and $\gamma_{IJ}$ the normalized overlap reduction function which encodes the baseline's geometry~\cite{PhysRevD.46.5250,Romano:2016dpx}. 
$S_0(f)$ is a normalization factor given by $S_0(f)=(9H_0^2)/(40\pi^2f^3)$ and $\Delta T$ is the time duration of the data used.\\
The energy density $\Omega_{\rm GW}(f)$ of the SGWB coming from unresolved CBC events is predicted to behave as a power law, with a slope of $\alpha = 2/3$, at lower frequencies where all the sources contribute in their inspiral phase~\cite{Regimbau:2011rp,sym14020270}. We assume the GW signal consists of equal parts tensorial plus- and cross-polarization, as expected by general relativity.
Given the normalization constant $S_0(f)$ contains a factor $f^{-3}$, this implies ${\rm{Re}}[\Tilde{s}^*_I(f)\Tilde{s}_J(f)] \propto f^{-7/3}$.

The PSD amplitude of the injected signal at the reference frequency $f_{\rm ref}$=\unit[25]{Hz} is $\sim 2.14 \times 10^{-49}$. This corresponds to $\Omega_{\rm GW} \sim 2.03 \times 10^{-8}$.
This amplitude is about two orders of magnitude larger than the value predicted by the LVK colaboration based on the results of their third observing run $2.1^{+2.9}_{-1.6} \times 10^{-10}$ \cite{KAGRA:2021kbb}. The amplitude is chosen such that the signal is clearly visible in a limited amount of data for illustrative purposes.

\subsubsection{Gaussian peaks}
\label{sec:datasets:GaussianPeaks}

Gaussian peaks in the frequency domain are used to simulate correlated noise sources. The Gaussian peaks are defined with the following PSD

\begin{equation}
    \label{eq:GPdefinition}
    S_n^{\rm GP}(f) = \left( \frac{A}{\sqrt{2\pi}}e^{\frac{-(f-\mu)^2}{2\sigma^2}} \right)^2,
\end{equation}
where the amplitude A, the frequency peak $\mu$ and its variance $\sigma$ are the free parameters. In our first example we use three peaks which have a peak frequency of $\mu$=\unit[10]{Hz}, $\mu$=\unit[50]{Hz} and $\mu$=\unit[90]{Hz}. Their respective amplitudes are $A^{\rm GP, 10Hz} = 4 \times 10^{-24}$, $A^{\rm GP, 50Hz} = 2 \times 10^{-24}$ and $A^{\rm GP, 90Hz} = 1.5 \times 10^{-24}$. For all three frequencies we use $\sigma=1$.

\subsubsection{Correlated magnetic noise}
\label{sec:datasets:Mag}

We use a magnetic noise spectrum described in \cite{PhysRevD.104.122006}. This low-frequency spectrum (\unit[$<100$]{Hz}) is based on measurements at the ET candidate site in the SoS Enattos mine in Sardegna, Italy \cite{Naticchioni:2020kfb,RRomero,Amann:2020jgo}. The measurements mainly focus on the Schumann resonances \cite{Schumann1,Schumann2}, which are electromagnetic excitation's in the cavity formed by the Earth's surface and the ionosphere. Due to their low frequency -- fundamental mode at $\sim$ \unit[7.8]{Hz} -- they are correlated over Earth-scale distances. 10\% of the measured magnetic spectrum at Sos Ennatos are used to get a clean measurement of the Schumann resonances. The amplitude of the fundamental mode is then rescaled to the amplitude of the 95\% to get a conservative estimate.

Due to the long distance correlations of the Schumann resonances it is safe to assume all interferometers of the ET be affected by this noise coherently. We thus assume this noise source to be identical, and fully correlated, among the three interferometers.

A final aspect one needs to understand is to which extent these ambient magnetic fields couple to the ET interferometers. In  \cite{PhysRevD.104.122006} the authors place upper limits on the maximal allowed coupling such that correlated magnetic noise would not impact SGWB searches. Here we assume a coupling function which is proportional to $f^{-2}$, as is measured at Virgo during the third observing run for frequencies below $\sim$ \unit[100]{Hz} \cite{galaxies8040082}. The amplitude is chosen to match the weekly-average of the magnetic coupling measured at Virgo \cite{galaxies8040082,PhysRevD.104.122006}. This yields an amplitude of \unit[$0.4 \times 10^{-10}$]{m/T} at \unit[100]{Hz}.

The power spectral density of this noise source will be referred to as $S_n^{\rm Mag}$ in the rest of the paper.

\subsubsection{Correlated Newtonian noise from body waves}
\label{sec:datasets:NN}

For the correlated Newtonian noise from body waves, we rely on \cite{PhysRevD.106.042008} which investigates how correlations in the seismic field and subsequent Newtonian noise from body waves has not only the potential to drastically impede SGWB searches but also limit the ET sensitivity. In this work, the authors also demonstrate that correlations over distance scales of couple of hundreds of metres can couple in five different ways to a given interferometer pair, e.g. $X$-$Y$. 

In this paper we assume the correlated Newtonian noise from body waves is only present in one vertex, which consists of the central station of the $X$ interferometer and the end station of the $Y$ interferometer. This implies that whereas in  \cite{PhysRevD.106.042008} a factor of 5 for the number of independent couplings is considered, we assume a factor of 2. This reduces the noise spectrum used in this paper by a factor of 2/5 with respect to the results presented in Fig. 13 of \cite{PhysRevD.106.042008} from which we extracted the median estimates.

The power spectral density of this noise source will be referred to as $S_n^{\rm NN}$ in the rest of the paper.

\subsubsection{Total data set}
\label{sec:datasets:total}

For the first example we use \unit[2000]{s} of data containing Gaussian noise colored with the design sensitivity of the xylophone configuration of the ET \footnote{This design is often also referred to as ET-D.}. In the rest of the paper we refer to this colored Gaussian noise as `ET noise' and use $S_n^{\rm ET}$\footnote{The PSD of the ET noise $S_n^{\rm ET}$ is not to be confused with the CSD of correlating the $T$ and $E$ channels, i.e. $S_n^{TE}$.} for its PSD. Note that despite the fact that we use identical noise, the $X$, $Y$ and $Z$ data are different Gaussian noise realisations.
On top of the detector noise we inject an artificially enhanced SGWB coming from CBC events, as introduced in Sec. \ref{sec:datasets:SGWB}.
Furthermore two Gaussian peaks, as described in Sec. \ref{sec:datasets:GaussianPeaks}, are injected in each interferometer. The injected peaks are correlated, such that every pair of interferometers has one correlated peak. More specifically we inject a Gaussian peak at \unit[10]{Hz} and \unit[50]{Hz} in the $X$ interferometer, at \unit[10]{Hz} and \unit[90]{Hz} in the $Y$ interferometer and at \unit[50]{Hz} and \unit[90]{Hz} in the $Z$ interferometer. 

If we refer to the PSD of the Gaussian peak at \unit[10]{Hz} by $S_n^{\rm GP, 10Hz}$, and equivalent for the other Gaussian peaks, we can write the PSD and CSD of the $X$, $Y$ and $Z$ interferometers as follow

\begin{equation}
\begin{aligned}
\label{eq:dataset-toyex}
    S_n^X(f) &=  S_n^{\rm ET}(f) + S_n^{\rm GP, 10Hz}(f) + S_n^{\rm GP, 50Hz}(f) \\
    S_n^Y(f) &=  S_n^{\rm ET}(f) + S_n^{\rm GP, 10Hz}(f) + S_n^{\rm GP, 90Hz}(f) \\
    S_n^Z(f) &=  S_n^{\rm ET}(f) + S_n^{\rm GP, 50Hz}(f) + S_n^{\rm GP, 90Hz}(f) \\
    S_n^{XY}(f) &= S_n^{\rm GP, 10Hz}(f)  \\
    S_n^{XZ}(f) &= S_n^{\rm GP, 50Hz}(f)  \\
    S_n^{YZ}(f) &= S_n^{\rm GP, 90Hz}(f). \\
\end{aligned}
\end{equation}

For the second example, we use \unit[2000]{s} of ET noise data on top of which we inject the same SGWB as in previous example.
We also add the correlated magnetic and Newtonian noise as respectively introduced in Sec. \ref{sec:datasets:Mag} and Sec. \ref{sec:datasets:NN}. The magnetic noise is correlated and identical among the $X$, $Y$ and $Z$ interferometers. The Newtonian noise is also correlated but only present in the $X$ and $Y$ interferometers, leading to non-identical noise sources in the three different ET interferometers.

With the introduced notations for the different spectral densities we can write the spectral densities as follow:

\begin{equation}
\begin{aligned}
\label{eq:dataset}
    S_n^X(f) &=  S_n^{\rm ET}(f) + S_n^{\rm Mag}(f) + S_n^{\rm NN}(f) \\
    S_n^Y(f) &=  S_n^{\rm ET}(f) + S_n^{\rm Mag}(f) + S_n^{\rm NN}(f) \\
    S_n^Z(f) &=  S_n^{\rm ET}(f) + S_n^{\rm Mag}(f) \\
    S_n^{XY}(f) &= S_n^{\rm Mag}(f) + S_n^{\rm NN}(f) \\
    S_n^{XZ}(f) &= S_n^{\rm Mag}(f)  \\
    S_n^{YZ}(f) &= S_n^{\rm Mag}(f). \\
\end{aligned}
\end{equation}

\subsection{Demonstration} 
\label{sec:demonstration:demo}

\subsubsection{Toy example with correlated Gaussian peaks}
\label{sec:demonstration:demo:example1}

\begin{figure*}[t]
    \includegraphics[width=0.8\textwidth]{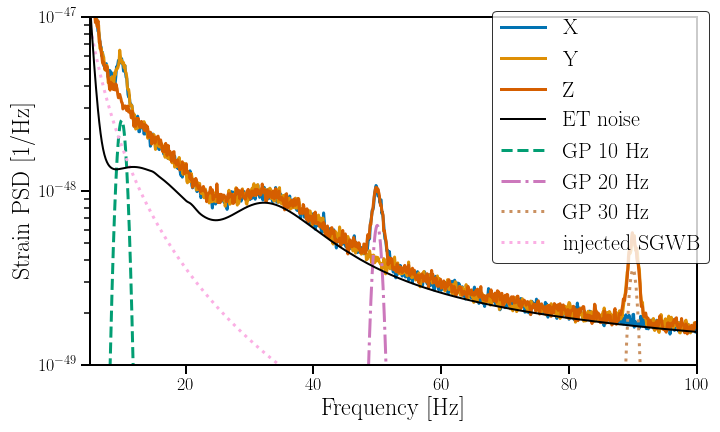} 
    \caption{The PSDs of the $X$, $Y$, $Z$ channels, the ET noise and the injected correlated noise $S_n^{\rm GP, 10}$, $S_n^{\rm GP, 50}$ and $S_n^{\rm GP, 90}$(see text for a more detailed description). The contribution of the injected SGWB $S_h^{\rm SGWB}$ is also shown.
    We point out that at high frequencies the $X$, $Y$ and $Z$ PSDs seem to not match the ET noise curve. This is due to the small but non-negligble contribution of the GW signal, as can be seen by the perfect match for the $T$ PSD in Fig. \ref{fig:AET_TXYZ_PSD}.}
    \label{fig:GP_XYZ_PSD}
\end{figure*}

Fig. \ref{fig:GP_XYZ_PSD} shows the PSD for the $X$, $Y$ and $Z$ interferometers. All the separate components are also shown as to compare their relative contributions. 
Due to the continuous presence of CBC signals it is impossible to directly estimate the noise spectral density from the observed spectrum. 

\begin{figure*}[t]
    \centering
    \includegraphics[width=0.49\textwidth]{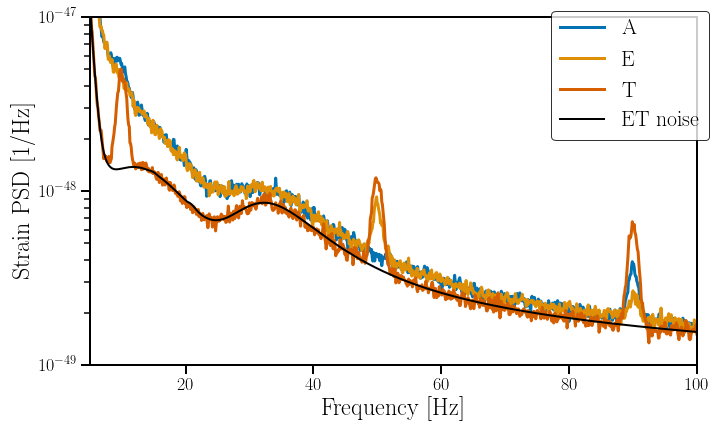} 
    \includegraphics[width=0.49\textwidth]{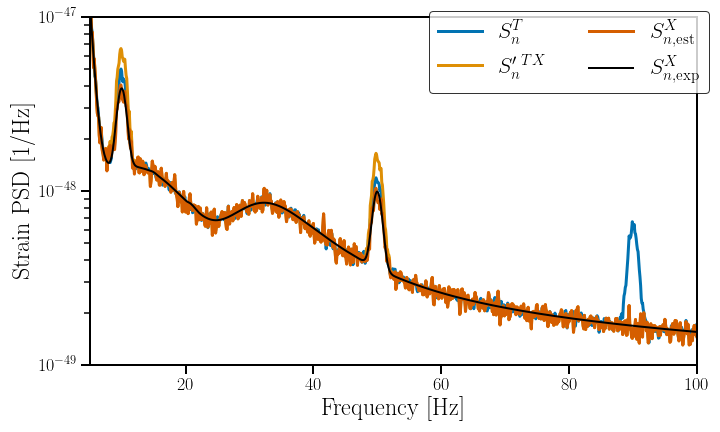}
    \includegraphics[width=0.49\textwidth]{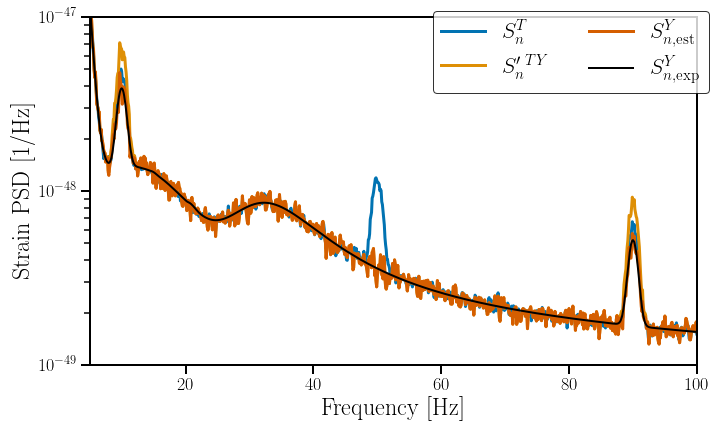} 
    \includegraphics[width=0.49\textwidth]{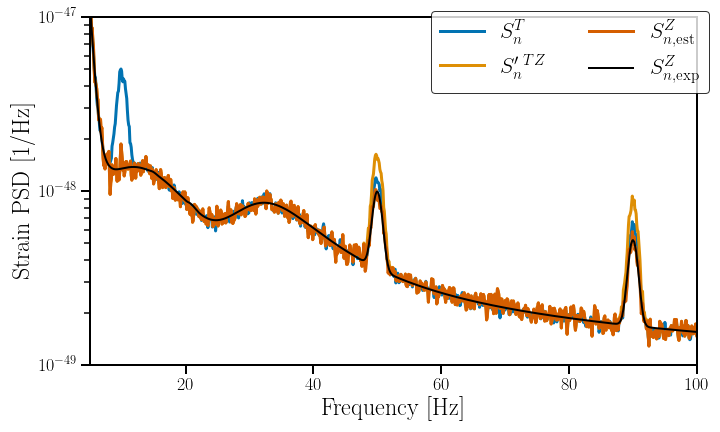} 
    \caption{Top left: The PSDs of the $A$, $E$, $T$ channels and the ET noise. Top right/bottom left/bottom right: The PSD of the null channel $T$, the CSD of the $T$ and $X$/$Y$/$Z$ channels, normalised such that it can serve as an estimate of $S_n^X$/$S_n^Y$/$S_n^Z$. The expected PSD of $X$/$Y$/$Z$, as shown in Eq. \ref{eq:dataset-toyex} and the estimated PSD as calculated in Eq. \ref{eq:CorrectedXPSDEstimation-toyex} and Eq. \ref{eq:CorrectedYZPSDEstimation-toyex}.}
    \label{fig:GP_AET_TXYZ_PSD}
\end{figure*}

The top left panel of Fig. \ref{fig:GP_AET_TXYZ_PSD} shows the spectral densities of the $A$, $E$ and $T$ channels. This figure illustrates that the null channel is indeed free of GWs. Therefore, as pointed out by \cite{PhysRevD.105.122007}, one could hope to use $S_{n}^{T}$ as an estimate for the noise PSD of e.g. the $X$ interferometer $S_{n}^{X}$, or alternatively, use $S_{n}^{\prime}{}^{TX}$ in case non-identical noise in the different $X$, $Y$ and $Z$ interferometers is suspected.
However as shown by the right panel of Fig. \ref{fig:GP_AET_TXYZ_PSD}, both $S_{n}^{T}$ and $S_{n}^{\prime}{}^{TX}$ are biased for our dataset. To be more specific two biases are present.

First, the PSD of the $T$ channel includes all noise features present in any of the 3 interferometers. Whereas the $X$ interferometer only contains the Gaussian peaks at \unit[10]{Hz} and \unit[50]{Hz}, $S_{n}^{T}$ includes all three Gaussian peaks. $S_{n}^{T}$ is thus a biased estimate of the $X$ interferometer. A similar statement is applicable to the two other interferometers, where respectively the \unit[50]{Hz} and \unit[10]{Hz} peak are not present in the $Y$ and $Z$ interferometers, but they are in $S_{n}^{T}$.

This bias can be addressed by using the cross-correlation between the null channel and the interferometer of interest, i.e. $S_{n}^{\prime}{}^{TX}$, as it has been first proposed in \cite{PhysRevD.105.122007}. However, this also yields a biased estimate since it does not take into account the fact that some of the noise sources are correlated. This is reflected in a general overestimate due to not properly taking the correlation into account. As shown in Eq. \ref{eq:differentSummary} and Eq. \ref{eq:differentSummary_TXYZ}, these non zero correlations imply that the cross-correlation between e.g. the $T$ and $X$ channel differs from the noise PSD $S_n^{X}$. Only by taking both elements, i.e. non-identical and correlated noise, one can build an unbiased PSD estimator for the different interferometers as explained below.

We now build an unbiased PSD estimator $S_{n,est}^X$ using Eq. \ref{eq:differentSummary_TXYZ} as follows

\begin{equation}
    \label{eq:CorrectedXPSDEstimation-toyex}
    \begin{aligned}
      S_{n}^{\prime}{}^{TX} & = \sqrt(3) * S_{n}^{TX}\\
      & = S_n^X +  S_n^{XY} +  S_n^{XZ} \\
      & = S_{n,{\rm est}}^X + S_n^{\rm GP, 10} + S_n^{\rm GP, 50} \\
      & \Downarrow \\
      S_{n,{\rm est}}^X & = S_{n}^{\prime}{}^{TX} - S_n^{\rm GP, 10} - S_n^{\rm GP, 50}.
    \end{aligned}
\end{equation}

Equivalently we find
\begin{equation}
    \label{eq:CorrectedYZPSDEstimation-toyex}
    \begin{aligned}
      S_{n,{\rm est}}^Y & = S_{n}^{\prime}{}^{TY} - S_n^{\rm GP, 10} - S_n^{\rm GP, 90} \\
      S_{n,{\rm est}}^Z & = S_{n}^{\prime}{}^{TZ} - S_n^{\rm GP, 50} - S_n^{\rm GP, 90}.
    \end{aligned}
\end{equation}

The spectral densities $S_{n}^{\prime}{}^{TX}$, $S_{n}^{\prime}{}^{TY}$ and $S_{n}^{\prime}{}^{TZ}$ can be extracted from the data. For this, we use the entire stretch of \unit[2000]{s} of data, split it in \unit[10]{s} long segments over which FFT's are calculated using 50\% overlapping Hann-windows. We average over the different segments. We assume to know perfectly $S_n^{\rm GP, 10}$, $S_n^{\rm GP, 50}$ and $S_n^{\rm GP, 90}$ that are subtracted from $S_{n}^{\prime}{}^{TX}$, $S_{n}^{\prime}{}^{TX}$ and $S_{n}^{\prime}{}^{TZ}$. In Sec. \ref{sec:Outlook} we discuss how this assumption can be dropped by developing a Bayesian estimation framework. 

The top right and bottom panels of Fig \ref{fig:GP_AET_TXYZ_PSD} compare the estimated PSDs to the expected ones for the three interferometers $X$, $Y$ and $Z$.
The expected PSDs $S_{n,{\rm exp}}$ are given by Eq. \ref{eq:dataset-toyex} where we have used the simulated data.
The agreement is excellent and much better than with the other estimators $S_n^T$ or $S_n^{\prime~TX}$, $S_n^{\prime~TY}$ and $S_n^{\prime~TZ}$.

\begin{figure*}[t]
    \centering
    \includegraphics[width=0.49\textwidth]{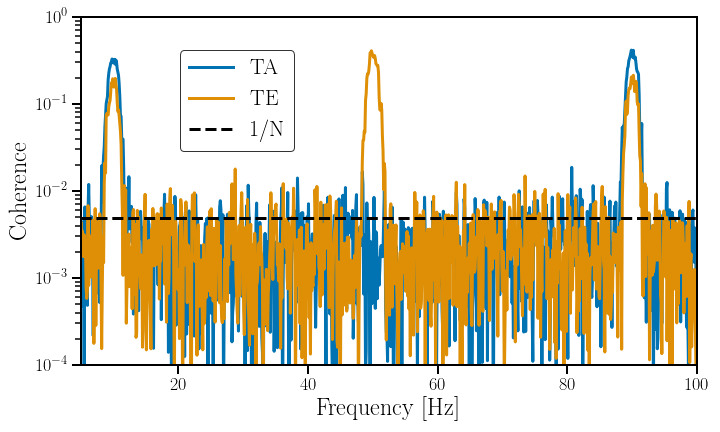} 
    \includegraphics[width=0.49\textwidth]{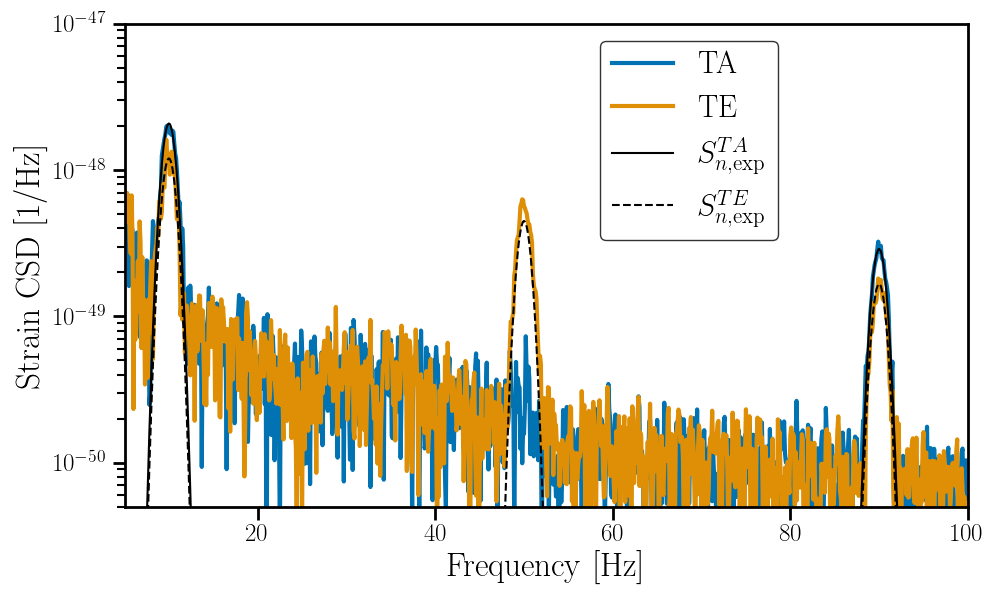} 
    \caption{Left: the coherence between the $T$ and $A$, $E$ channels. The black dashed line represents the level of coherence expected from independent Gaussian data, which goes approximately as $1/N$, where $N$ is the number of time segments over which was averaged.
    Right: the modulus of the CSD between the $T$ and $A$, $E$ channels.
    The expected cross spectral densities associated with the $T$ and $A$ channels, and $T$ and $E$ channels given by Eq.~\ref{eq:TATE_expected} for the toy model example are shown in black. The expected CSD is in agreement with the observed CSD.}
    \label{fig:AETCohCSD}
\end{figure*}

As mentioned in Sec. \ref{sec:nullchannel_complex}, the cross correlation between the $T$ and $A$, $E$ channels ($S_n^{TA}$ and $S_n^{TE}$) contains additional information on the non-identical character of the correlated noise. The left panel of Fig. \ref{fig:AETCohCSD} shows the coherence between the $T$ channel and the $A$ and $E$ channels. This coherence is consistent with a Gaussian noise origin \footnote{The expected coherence for uncorrelated data goes approximately as 1/N, where N is the number of time segments over which the coherence is averaged.} apart from the Gaussian peaks. This indicates, correctly, that the only part of the noise in the $X$, $Y$ and $Z$ channels which is not identical are the Gaussian peaks. The right panel of Fig. \ref{fig:AETCohCSD} shows the modulus of the cross spectral densities estimated from the data on top of the expected quantities which are derived in Eq. \ref{eq:TATE_expected}.
\begin{widetext}
\begin{equation}
\label{eq:TATE_expected}
    \begin{aligned}
      |S_{n, {\rm exp}}^{TA}| =  & \frac{1}{\sqrt{6}} | S_n^Z(f) - S_n^X(f) + S_n^{YZ}(f) - S_n^{XY}(f) |  \\
      \Downarrow & \text{   Eq. \ref{eq:dataset-toyex}}\\
      =& \frac{1}{\sqrt{6}} |S_n^{\rm ET} + S_n^{\rm GP, 50} + S_n^{\rm GP, 90} - (S_n^{\rm ET} + S_n^{\rm GP, 10} + S_n^{\rm GP, 50} ) + S_n^{\rm GP, 90} - S_n^{\rm GP, 10}|\\
      =& \frac{2}{\sqrt{6}} | - S_n^{\rm GP, 10} + S_n^{\rm GP, 90}|\\
      =& \frac{2}{\sqrt{6}} \sqrt{\mathcal{R}e\left( - S_n^{\rm GP, 10} + S_n^{\rm GP, 90}\right)^2 + \mathcal{I}m \left(   - S_n^{\rm GP, 10} + S_n^{\rm GP, 90} \right)^2}\\      
      \Downarrow & \text{  All the quantities involved are real;  }\left( A - B\right)^2 = \left( A + B\right)^2 -4AB\\
      =& \frac{2}{\sqrt{6}} \sqrt{\left(S_n^{\rm GP, 10} + S_n^{\rm GP, 90}\right)^2-4S_n^{\rm GP, 10}S_n^{\rm GP, 90}}\\
      \Downarrow & \text{  } S_n^{\rm GP, i}S_n^{\rm GP, j} \ll \left(S_n^{\rm GP, k}\right)^2 \text{, where  }i\neq j \text{ and $i$,$j$,$k$ $\in$ [`10',`50',`90']} \\
      \approx & \frac{2}{\sqrt{6}} \left(S_n^{\rm GP, 10} + S_n^{\rm GP, 90}\right)\\
    |S_{n, {\rm exp}}^{TE}| =  & \frac{1}{3\sqrt{2}} | S_n^X(f) - 2 S_n^Y(f) + S_n^Z(f) - S_n^{XY}(f) + 2 S_n^{XZ}(f) - S_n^{YZ}(f)  |  \\  
    =  & \frac{1}{3\sqrt{2}} | S_n^{\rm ET} + S_n^{\rm GP, 10} + S_n^{\rm GP, 50} - 2 (S_n^{\rm ET} + S_n^{\rm GP, 10} + S_n^{\rm GP, 90} ) \\
    & \text{\phantom{$\frac{1}{3\sqrt{2}} |$}} + S_n^{\rm ET} + S_n^{\rm GP, 50} + S_n^{\rm GP, 90}  - S_n^{\rm GP, 10} + 2 S_n^{\rm GP, 50} - S_n^{\rm GP, 90} |  \\  
    =  & \frac{2}{3\sqrt{2}} |- S_n^{\rm GP, 10} + 2S_n^{\rm GP, 50} - S_n^{\rm GP, 90} |  \\  
    =& \frac{2}{3\sqrt{2}} \sqrt{\mathcal{R}e\left(- S_n^{\rm GP, 10} + 2 S_n^{\rm GP, 50} -  S_n^{\rm GP, 90}  \right)^2 + \mathcal{I}m \left( -  S_n^{\rm GP, 10} + 2 S_n^{\rm GP, 50} -  S_n^{\rm GP, 90}  \right)^2}\\   
    \Downarrow & \text{  All the quantities involved are real;   }\left( A - B\right)^2 = \left( A + B\right)^2 -4AB\\
    =& \frac{2}{3\sqrt{2}} \sqrt{\left( S_n^{\rm GP, 10} + 2 S_n^{\rm GP, 50} +  S_n^{\rm GP, 90}  \right)^2 - 8 S_n^{\rm GP, 50} \left( S_n^{\rm GP, 10} + S_n^{\rm GP, 90}\right)}\\    
     \Downarrow & \text{  } S_n^{\rm GP, i}S_n^{\rm GP, j} \ll \left(S_n^{\rm GP, k}\right)^2 \text{, where  }i\neq j \text{ and $i$,$j$,$k$ $\in$ [`10',`50',`90']}  \\
    =  & \frac{2}{3\sqrt{2}} ( S_n^{\rm GP, 10} + 2 S_n^{\rm GP, 50} + S_n^{\rm GP, 90} ),  \\  
    \end{aligned}
\end{equation}
\end{widetext}
As one can see in the right panel of Fig. \ref{fig:AETCohCSD} the expected quantities calculated in Eq. \ref{eq:TATE_expected} are consistent with CSD estimated from the data. 

\subsubsection{Superposition of SGWB, correlated magnetic and correlated Newtonian noise}
\label{sec:demonstration:demo:example2}

\begin{figure*}[t]
    \centering
    \includegraphics[width=0.8\textwidth]{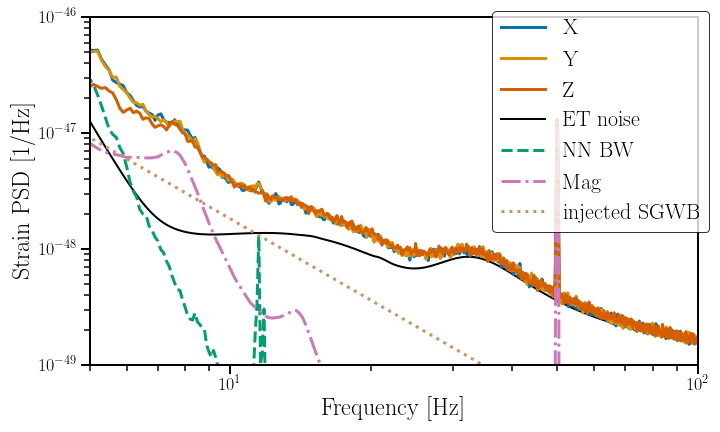} 
    \caption{The PSDs of the $X$, $Y$, $Z$ channels, the ET noise and the injected correlated Newtonian noise (NN) and magnetic noise (Mag). The contribution of the injected SGWB $S_h^{\rm SGWB}$ is also shown.}
    \label{fig:XYZ_PSD}
\end{figure*}

Fig. \ref{fig:XYZ_PSD} shows the power spectral density spectrum for the $X$, $Y$ and $Z$ interferometers with the second data-set composed of Gaussian noise on top of which correlated magnetic, Newtonian noise and SGWB signal have been added. All the separate components are shown as to compare their relative contributions.

\begin{figure*}[t]
    \centering
    \includegraphics[width=0.49\textwidth]{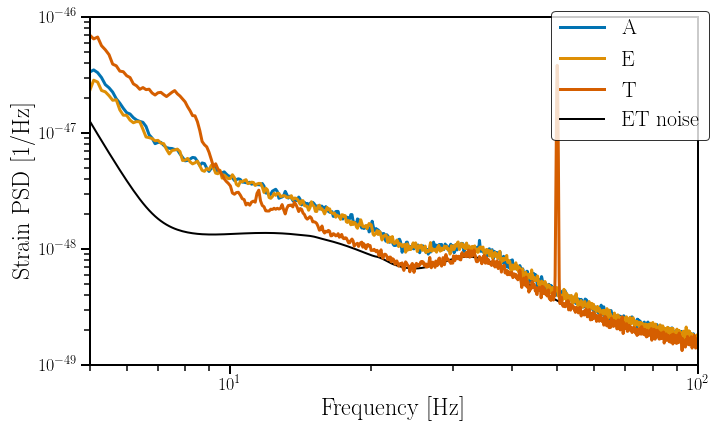} 
    \includegraphics[width=0.49\textwidth]{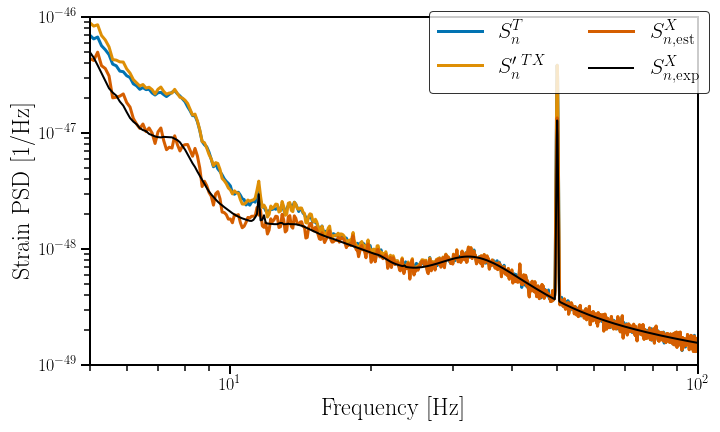}
    \includegraphics[width=0.49\textwidth]{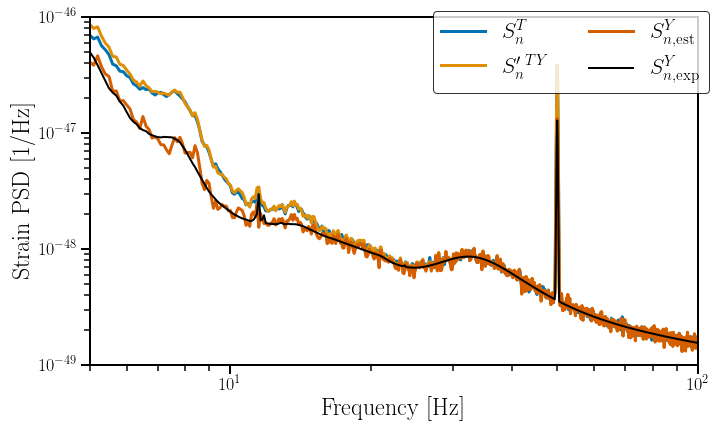} 
    \includegraphics[width=0.49\textwidth]{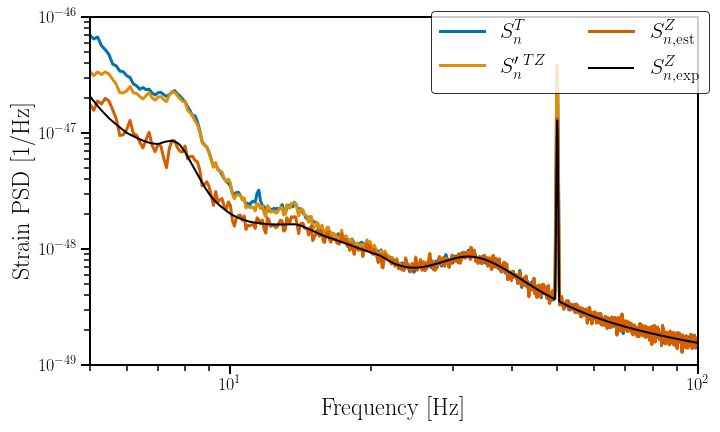} 
    \caption{Top left: The PSDs of the $A$, $E$, $T$ channels and the ET noise. Top right/bottom left/bottom right: The PSD of the null channel $T$, the CSD of the $T$ and $X$/$Y$/$Z$ channels, normalised such that it can serve as an estimate of $S_n^X$/$S_n^Y$/$S_n^Z$. The expected PSD of $X$/$Y$/$Z$, as shown in Eq. \ref{eq:dataset} and the estimated PSD as calculated in Eq. \ref{eq:CorrectedXYZPSDEstimation}. As explained in the text, $S_n^T$ and $S_{n}^{\prime}{}^{TX}$/$S_{n}^{\prime}{}^{TY}$/$S_{n}^{\prime}{}^{TZ}$ yield a biased PSD estimate and one needs to use $S_{n,{\rm est}}^X$/$S_{n,{\rm est}}^Y$/$S_{n,{\rm est}}^Z$, derived in Eq. \ref{eq:CorrectedXYZPSDEstimation} using Eq. \ref{eq:differentSummary_TXYZ}.}
    \label{fig:AET_TXYZ_PSD}
\end{figure*}

The top left panel of Fig. \ref{fig:AET_TXYZ_PSD} shows the spectral densities of the $A$, $E$ and $T$ channels. Compared to the previous example it is slightly more difficult to immediately see the $T$ channel is indeed free of GW due to the presences of the magnetic and Newtonian noise. However, based on Fig. \ref{fig:XYZ_PSD} one could expect negligible effect from either the magnetic or Newtonian noise above \unit[20]{Hz}, apart from the magnetic resonance at \unit[50]{Hz}.

In the previous example we showed how $S_{n}^{T}$ as well as $S_{n}^{\prime}{}^{TX}$ yield a biased 
estimate for the noise PSD of the $X$ interferometer $S_{n}^{X}$, due to the presence of non-identical and correlated noise. Afterwards we showed one can build unbiased estimators of the interferometers' PSD using Eq. \ref{eq:differentSummary} and Eq. \ref{eq:differentSummary_TXYZ}.
We now provide another demonstration with a more realistic scenario where correlated magnetic and Newtonian noise are present.

Similar to the previous example we compare our estimators with the expected quantities given by Eq. \ref{eq:dataset}. The estimated PSDs for this example are given by: 

\begin{equation}
    \label{eq:CorrectedXYZPSDEstimation}
    \begin{aligned}
      S_{n,{\rm est}}^X & = S_{n}^{\prime}{}^{TX} - 2 * S_n^{\rm Mag} - S_n^{\rm NN} \\
      S_{n,{\rm est}}^Y & = S_{n}^{\prime}{}^{TY} - 2 * S_n^{\rm Mag} - S_n^{\rm NN} \\
      S_{n,{\rm est}}^Z & = S_{n}^{\prime}{}^{TZ} - 2 * S_n^{\rm Mag}.
    \end{aligned}
\end{equation}

$S_n^{\rm Mag}$ and $S_n^{\rm NN}$ can be measured for instance using witness sensors. In a more realistic scenario the knowledge of each individual noise sources might be inaccurate because of inefficiencies introduced by the witness sensors. This source of inaccuracy should be more thoroughly investigated in future work.

The top right and two bottom panels of Fig \ref{fig:AET_TXYZ_PSD} show a good agreement between $S_{n,{\rm exp}}$ and $S_{n,{\rm est}}$ for respectively the $X$, $Y$ and $Z$ interferometers.

The previous two examples have shown how one can build unbiased PSD estimators based on the $T$ channel and an estimate of the correlated noise present in the data.
In the next section we discuss the next steps to develop a PSD estimation framework.

%%%%%%%%%%%%%
\section{Recipe to transform the null channel formalism into a PSD estimation framework}
\label{sec:Outlook}
%%%%%%%%%%%%%

In previous sections, we have shown how one can construct the sky location independent null channel for an equilateral triangular GW interferometer configuration in a situation where there is correlated noise unique to each detector pair, as well as where all the interferometers have different noise levels. However, the examples assumed one has perfect knowledge of the correlated and individual noise sources. In the example one could get an estimation of $S_n^{\rm Mag}$ ($S_n^{\rm NN}$) from e.g. witness sensors measuring the magnetic (seismic) fields, combined with a measurement of the magnetic coupling function (formula to predict NN based on seismic fields).
In a real scenario for the ET, this knowledge will never be perfect. It is furthermore very likely some part of the noise is not properly understood. In this section, we propose further investigations so that one can estimate the noise PSDs as well as correlated noise levels for ET (or LISA) to observe GWs.

After the proof of concept of the $A$, $E$ and $T$ channels for the ET in earlier sections, one should develop a Bayesian framework to enable parameter estimation of the noise PSDs as well as correlated noise PSDs, that is $S_n^{I}$ and $S_n^{IJ}$ or in the more general scenario $S_n^{X}$, $S_n^{Y}$, $S_n^{Z}$, $S_n^{XY}$, $S_n^{XZ}$ and $S_n^{YZ}$ \cite{PhysRevD.92.064011,Christensen:2022bxb}. Here one should consider multiple scenarios such as identical noise sources, unique noise sources for each $X$, $Y$ and $Z$ interferometer, absence/presence of correlated noise. 
In the case of non-identical noise sources for the different interferometers it is beneficial to use the cross-correlation of the null channel $T$ with the $X$, $Y$ and $Z$ channels ($S_s^{TX}(f)$, $S_s^{TY}(f)$ and  $S_s^{TZ}(f)$) rather than using the PSD of $T$ ($S_s^{T}(f)$) to estimate the noise sources in the different interferometers. Alternatively one can also extract this information on the non-identical noise components in the $X$, $Y$ and $Z$ channels by using the cross spectral density of the null channel with the A and E channels ($S_s^{TA}(f)$ and  $S_s^{TE}(f)$), together with the null channel ($S_s^T$).
Furthermore, one should understand the effect of knowledge on (un)correlated noise sources, from possible witness sensors observing the noise sources (e.g. magnetometers, seismometeres, etc.), and how this can improve the final parameter estimation of the noise PSDs. 
Given the high dimensional problem, estimations of all the parameters with the desired accuracy might prove difficult in the most general scenario and more work is needed to achieve such a goal. To this extent, one should also take the non-stationary character of real data into account. This implies it is not trivial to just extend the duration of the data used to compute the different noise contributions using the null channel formalism, presented in this paper.

Third and finally, the knowledge on correlated noise sources and their properties should be understood and studied further if one wants to simulate a realistic scenario for the above techniques. In the context of the search for a SGWB using Earth-based interferometers there have already been significant investigations concerning the effects of correlated magnetic fields, e.g. Schumann resonances \cite{Thrane:2013npa,Thrane:2014yza,Coughlin:2016vor,Himemoto:2017gnw,Coughlin:2018tjc,Himemoto:2019iwd,Meyers:2020qrb}, also in the context of ET \cite{PhysRevD.104.122006}. 
A recent paper studies correlated seismic and Newtonian noise over distances of several hundreds of meters\footnote{\unit[300--500]{m} is the expected distance between the terminal and central stations of two different ET interferometers \cite{ETdesignRep}.} and their impact on stochastic searches using the ET \cite{PhysRevD.106.042008}. These results form the ideal starting point for studying correlated noise in the formalism described in this paper but more dedicated follow-up studies could look into the effect from earthquakes, the possibility of correlated seismic fields on the scale of \unit[10]{km} and the interplay between different coupling locations and to which extent there is constructive or destructive interference between e.g. correlated magnetic fields coupling to both input and end mirrors of the interferometers.

For LISA noise correlation studies resulting from the spatial environment can benefit from more detailed noise characterisation. These noises include for example the effect of micro-thrusters, the magnetic field and the temperature variations on the test-mass system~\cite{2022arXiv220403867B}. 

These research topics pose fundamental questions for understanding the data analysis environment at the ET and LISA, and should be studied in significantly more details over the course of the coming years before the instruments become operational to ensure they can fulfill their scientific goals. This is especially true for correlated noise, which would seriously affect the search for a SGWB.

%%%%%%%%%%%%%
\section{Conclusion}
\label{sec:Conclusion}
%%%%%%%%%%%%%

Future GW interferometers such as ET, CE and LISA will have sensitivities allowing them to resolve many transient sources to such an extent that they are expected to observe a large amount of overlapping signals \cite{PhysRevD.86.122001,PhysRevD.79.062002,2017arXiv170200786A}. This makes it difficult to estimate the noise PSD of the interferometers needed to perform GW searches.

A first method is to simultaneously model noise and signals using for instance a joint Bayesian parameter estimation framework~\cite{PhysRevD.92.064011,Christensen:2022bxb}. This paper is investigating a different approach relying on the sky location independent null channel which can be constructed for triangular configurations of interferometers ET and LISA. This sky location independent null channel is insensitive to GWs from any direction.

We introduce a formalism which is able to not only address correlated noise between the different interferometers, but also allow for non-identical noise both correlated and uncorrelated. The formalism goes beyond using the sky location independent null channel and relies on three linear combinations $A$, $E$ and $T$ of the three interferometers $X$, $Y$ and $Z$, which have been used in LISA before \cite{Tinto2005, PhysRevD.100.104055}. The advantage of relying on the $A$, $E$ and $T$ channels, of which the $T$ channel is a normalized version of the null channel, is that their cross-correlation spectra contain additional information in the case of non-identical noise sources. 
Alternatively one can correlate the null channel with the $X$, $Y$ and $Z$ channels which allows to estimation of the noise PSDs and CSDs in the case of non-identical noise compared by disentangling the non-identical noise . The later was also investigated in \cite{PhysRevD.105.122007}, however although mentioning the possible bias coming from correlated noise, the authors did not take it into account.

We illustrate the formalism considering simplified but realistic noise realisation for the ET on top of which a SGWB signal is added. 
We show the T channel is indeed insensitive to gravitational waves. 
Furthermore, we show that our formalism enables an unbiased estimation of the noise spectral densities of the $X$, $Y$ and $Z$ channels, even in the presence of correlated and non-identical noise.
These examples use knowledge of the noise sources to prove the mathematical consistency of this formalism, which in future research should be used as ingredient for a Bayesian estimation framework, as discussed in Sec. \ref{sec:Outlook}.

\acknowledgements

The authors acknowledge access to computational resources provided by the LIGO Laboratory supported by National Science Foundation Grants PHY-0757058 and PHY-0823459. GB thanks the laboratory  Artemis, Observatoire de la C\^ote d'Azur, for hospitality and welcome.

Furthermore, the authors would like to thank Q. Baghi, B. Goncharov, S. Shah and O. Hartwig for useful comments.

This paper has been given LIGO DCC number P2200126, Virgo TDS number VIR-0443A-22 and ET TDS number ET-0066A-22.

K.J. is supported by FWO-Vlaanderen via grant number 11C5720N.

%%%%%%%%%%%%%%%%%
\bibliography{references}
%%%%%%%%%%%%%%%%%

\appendix
%%%%%%%%%%%%%
\section{Antenna pattern functions for tensorial gravitational waves}
\label{sec:Appendix1-T}
%%%%%%%%%%%%%

We follow the description of the ET's triangle of the Fig. 1 from \cite{PhysRevD.86.122001} but interferometer indices 1,2,3 become X,Y,Z.

We introduce the orthogonal triad in the transverse trace less form as $(\mathbf{e_x}, \mathbf{e_y}, \mathbf{e_z})$, which allows us to write the basis polarization tensors : 

\begin{equation}
    \begin{aligned} 
\mathbf{e}_+ &= \mathbf{e_x} \otimes \mathbf{e_x} - \mathbf{e_y} \otimes \mathbf{e_y} \\ 
\mathbf{e}_{\times} &= \mathbf{e_x} \otimes \mathbf{e_y} + \mathbf{e_y} \otimes \mathbf{e_x}
\end{aligned}
\end{equation}

with $\mathbf{h} = h_+\mathbf{e}_+ +  h_+\mathbf{e}_{\times}$ and $h_+, h_{\times}$ the tensor plus-, respectively cross-polarization components. 

The symmetric trace-free tensors representing the three ET interferometers are given by: 

\begin{equation}
    \begin{aligned} 
{d}^X &= \frac{1}{2}\mathbf{e_1} \otimes \mathbf{e_1} - \mathbf{e_2} \otimes \mathbf{e_2} \\ 
{d}^Y &= \frac{1}{2}\mathbf{e_2} \otimes \mathbf{e_2} - \mathbf{e_3} \otimes \mathbf{e_3} \\ 
{d}^Z &= \frac{1}{2}\mathbf{e_3} \otimes \mathbf{e_3} - \mathbf{e_1} \otimes \mathbf{e_1} \\ 
\end{aligned}
\end{equation}
with the basis of the three arms of the ET's configuration given by $\mathbf{e}_1;\mathbf{e}_2;\mathbf{e}_3 = \frac{1}{2}(\sqrt{3},-1,0);  \frac{1}{2}(\sqrt{3},1,0); (0,1,0)$. 

For each interferometer ($I = X,Y,Z$), the interferometer response $h^I(f)$ is given by the product between the detector tensor $\mathbf{d}^I$ and the tensor $\mathbf{h}$: 

\begin{equation}
\begin{aligned}
      h^I(t) &= d^I_{ij}h^{ij} = d^I_{ij}e_+^{ij}h_+ + d^I_{ij}e_{\times}^{ij}h_{\times} \\ 
      &= F_+^I h_+ + F_+^{\times} h_{\times}
\end{aligned}
\end{equation}
with $F^I_p$ the antenna pattern function of interferometer $I$ for a GW with polarization $p$.

According to ~\cite{PhysRevD.86.122001}, it is possible to write the unit vector of the basis ($xyz$) in the radiation frame. We introduce the polarization angle $\psi$ as $\cos \psi = \mathbf{e}_{\theta} \cdot \mathbf{e}_x$, so, we can write the antenna pattern function as a function of the sky position $(\theta, \phi)$. For example, the antenna pattern function of the interferometer $X$ is :

\begin{equation}
    \begin{aligned} 
F_{+}^X &= \frac{-\sqrt{3}}{4}\Big[\big( 1 + \cos^2 \theta  \big) \sin2\phi \cos2\psi + 2 \cos\theta \cos2\phi \sin2\psi \Big] \\
&= \alpha(\theta,\phi)\cos2\psi + \beta(\theta,\phi)\sin2\psi \\ 
F_{\times}^X &= \frac{\sqrt{3}}{4}\Big[\big( 1 + \cos^2 \theta  \big) \sin2\phi \sin2\psi - 2 \cos\theta \cos2\phi \cos2\psi \Big] \\ 
&= -\alpha(\theta,\phi)\sin2\psi + \beta(\theta,\phi)\cos2\psi ,
\end{aligned}
\end{equation}

where we have defined $\alpha(\theta,\phi) = -\frac{\sqrt{3}}{4} \big( 1 + \cos^2 \theta  \big) \sin2\phi$ and $\beta(\theta,\phi) =  -\frac{\sqrt{3}}{2} \cos\theta \cos2\phi$. It is possible, given the equilateral triangle configuration, to calculate the antenna pattern function of the interferometer $Y$ and $Z$ as a rotation of $2\pi/3$. This is given by changing the angle $\phi$ such that $\phi \longrightarrow \phi +\frac{2\pi}{3}$ for $Y$ arm and $\phi \longrightarrow \phi -\frac{2\pi}{3}$ for $Z$ arm.

\begin{equation}
    \begin{aligned} 
&F_{p}^Y(\theta, \phi, \psi) = F_{p}^X(\theta, \phi+\frac{2\pi}{3}, \psi) \\ 
&F_{p}^Z(\theta, \phi, \psi) = F_{p}^X(\theta, \phi-\frac{2\pi}{3}, \psi) \\ 
\end{aligned}
\end{equation}
with the polarization, $p = +, \times$. 
We can also write $\sum_p (F^X_p)^2$:

\begin{equation}
    \begin{aligned} 
(F_{+}^X)^2 +& (F_{\times}^X)^2   =  (\alpha\cos2\psi + \beta\sin2\psi)^2 + \\
&(-\alpha\sin2\psi + \beta\cos2\psi)^2   \\ 
=& \alpha^2 + \beta^2 \\
=& \frac{3}{16} \Big[\big( 1 + \cos^2 \theta  \big)^2 \sin^2 2\phi  + 4 \cos^2\theta \cos^2 2\phi  \Big]
\end{aligned}
\end{equation}

We assume that the noise and GWs are not correlated. The PSD of the $X$ channel is given by

\begin{equation}\label{eq:XnoCorr}
   \begin{aligned}
   \langle X(f)& X^*(f^{\prime})\rangle  = \left< n^X(f) n^{X*}(f^{\prime})\right> + \left< d^X_{ij} h^{ij}(f) (d^X_{ij}h^{ij}(f^{\prime}))^*\right>  \\
   & =\frac{1}{2}\delta(f-f^{\prime})S_n^X + \left\langle \sum_p F^X_p h_p \left(\sum_{p\prime}  F^X_{p\prime} h_{p\prime}\right)^* \right\rangle  \\   
   & = \frac{1}{2}\delta(f-f^{\prime})  S_n^X + \left\langle \sum_p  \left (F_p^X h_p\right) ^2  \right\rangle ,  \\
   & = \frac{1}{2}\delta(f-f^{\prime}) \left[ S_n^X + \sum _p S_{h_p} \int_{sky} \left (F_p^X\right) ^2 \right],  \\
   \end{aligned} 
\end{equation}
where $S_X$ is the noise PSD of interferometer $X$ and $S_{h_p}$ the PSD of the GW signal for an isotropic SGWB present with polarization $p$.

According to Eq.~\ref{eq:XnoCorr}, we can calculate the antenna patterns over the sky : 

\begin{equation}
    \begin{aligned}
    \int_{sky} (F_+^X)^2 + (F_{\times}^X)^2 &= \frac{1}{4\pi^2}\int_{0}^{\pi} \int_{0}^{2\pi} \int_{0}^{\pi} \sin\theta\mathrm{d}\theta \mathrm{d}\phi \mathrm{d}\psi \Big[(F_+^X)^2 \\
    &+ (F_{\times}^X)^2 \Big]      = \frac{1}{4\pi^2} \frac{6\pi^2}{5} = \frac{3}{10}
\end{aligned}
\end{equation}

The calculation for the $Y$ and $Z$ arms is identical and redundant. We can thus write : 

\begin{equation}
\begin{aligned}
    \int_{sky} (F_+^X)^2 + (F_{\times}^X)^2 = \int_{sky} (F_+^Y)^2 + (F_{\times}^Y)^2 & \\
    = \int_{sky} (F_+^Z)^2 + (F_{\times}^Z)^2
      = \frac{3}{10}
      \end{aligned}
\end{equation}

It is also possible to calculate the cross arm integration of the pattern antenna. We can notice first that we can write : 

\begin{equation}
    \begin{aligned} 
& (F_{+}^XF_{+}^Y) + (F_{\times}^XF_{\times}^Y) =  (\alpha\cos2\psi + \beta\sin2\psi)(\alpha'\cos2\psi + \beta'\sin2\psi) \\
&+ (-\alpha\cos2\psi + \beta\sin2\psi)(-\alpha'\cos2\psi + \beta'\sin2\psi) \\ 
&= \alpha\alpha' + \beta\beta' = \frac{3}{16} \Big[\big( 1 + \cos^2 \theta  \big)^2 \sin 2\phi \sin 2\phi'  \\
&+ 4 \cos^2\theta \cos2\phi \cos2\phi' \Big]
\end{aligned}
\end{equation}
with $\alpha' = \alpha(\theta,\phi+\frac{2\pi}{3})$, $\beta' = \beta(\theta,\phi+\frac{2\pi}{3})$ and $\phi' = \phi+\frac{2\pi}{3}$

we can calculate the antenna patterns over the sky : 

\begin{equation}
    \begin{aligned}
    &\int_{sky} (F_{\times}^XF_{\times}^Y) + (F_{+}^XF_{+}^Y)  \\ 
    &= \frac{1}{4\pi^2}\int_{0}^{\pi} \int_{0}^{2\pi} \int_{0}^{\pi}\sin\theta\mathrm{d}\theta \mathrm{d}\phi \mathrm{d}\psi \Big[(F_{\times}^XF_{\times}^Y) + (F_{+}^XF_{+}^Y) \Big] \\
    & = \frac{1}{4\pi^2} \frac{-3\pi^2}{5} = -\frac{3}{20}
    \end{aligned}
\end{equation}

The calculation for the $(XZ)$ and $(YZ)$ arms is also identical and redundant. We can thus write : 

\begin{equation}
    \begin{aligned}
    &\int_{sky} (F_{\times}^XF_{\times}^Y) + (F_{+}^XF_{+}^Y)= \int_{sky} (F_{\times}^XF_{\times}^Z)+ (F_{+}^XF_{+}^Z) \\
    &  = \int_{sky} (F_{\times}^YF_{\times}^Z) + (F_{+}^YF_{+}^Z)
      = -\frac{3}{20}
    \end{aligned}
\end{equation}

We can summarize the contribution of an isotropic SGWB to each channel according to the Eq.~\ref{eq:differentSummary}: 

\begin{equation}
\begin{aligned}
  \left<X(f)X^*(f')\right>&= \frac{1}{2}\delta(f-f')\left[ S_n^X(f) + \frac{3}{10}S_h(f)\right] \\
     \left<A(f)A^*(f')\right>&= \frac{1}{2}\delta(f-f')\left[ S_n^A(f) + \frac{9}{20}S_h(f)\right]  \\  
     \left<E(f)E^*(f')\right>&= \frac{1}{2}\delta(f-f')\left[ S_n^E(f)+ \frac{9}{20}S_h(f)\right]  \\  
     \left<A(f)E^*(f')\right>&= \frac{1}{2}\delta(f-f')\left[ S_n^{AE}(f) + 0S_h(f)\right]  \\  
\end{aligned}
\end{equation}

%%%%%%%%%%%%%
\section{Antenna pattern functions for non-GR polarization gravitational waves}
\label{sec:Appendix2-SV}
%%%%%%%%%%%%%
The non-GR polarization can also be defined from the orthogonal triad $(\mathbf{e_x}, \mathbf{e_y}, \mathbf{e_z})$ \cite{PhysRevLett.120.201102}. We define two other kind of polarization, the vector $(x,y)$ and scalar $(b,l)$ modes \citep{PhysRevD.105.064033,PhysRevD.100.042001}. 
\begin{equation}
    \begin{aligned} 
\mathbf{e}_x &= \mathbf{e_x} \otimes \mathbf{e_z}+ \mathbf{e_z} \otimes \mathbf{e_x} \\ 
\mathbf{e}_y &= \mathbf{e_y} \otimes \mathbf{e_z} + \mathbf{e_z} \otimes \mathbf{e_y}
\end{aligned}
\end{equation}

and

\begin{equation}
    \begin{aligned} 
\mathbf{e}_b &= \mathbf{e_x} \otimes \mathbf{e_x}+ \mathbf{e_y} \otimes \mathbf{e_y} \\ 
\mathbf{e}_l &= \mathbf{e_z} \otimes  \mathbf{e_z}
\end{aligned}
\end{equation}

For interferometer $X$, the antenna pattern of the non-GR modes are given by,

\begin{itemize}
    
    \item Vector modes:
        \begin{equation}
            \begin{aligned}
                F^x(\theta, \phi, \psi)&=\frac{-\sqrt{3}}{2}\sin\theta\left[\cos\theta\cos\psi\sin2\phi+\sin\psi\cos2\phi \right] \\
                F^y(\theta, \phi, \psi)&=\frac{\sqrt{3}}{2}\sin\theta\left[\cos\theta\sin\psi\sin2\phi-\cos\psi\cos2\phi \right] \\
            \end{aligned}
        \end{equation}
    \item Scalar modes:
        \begin{equation}
            \begin{aligned}
                F^b(\theta, \phi)&=\frac{\sqrt{3}}{4}\sin^2\theta\sin2\phi \\
                 F^l(\theta, \phi)&=\frac{-\sqrt{3}}{4}\sin^2\theta\sin2\phi \\
            \end{aligned}
        \end{equation}    
\end{itemize}

The integration over the sky for differences configuration are :
\begin{itemize}
    
    \item Vector modes:
        \begin{equation}
        \begin{aligned}
                &\int_{sky} (F_x^X)^2 + (F_y^X)^2 = \int_{sky} (F_x^Y)^2 + (F_y^Y)^2 \\&= \int_{sky} (F_x^Z)^2 + (F_y^Z)^2
      = \frac{3}{10} \\
               & \int_{sky} (F_y^XF_y^Y) + (F_x^XF_x^Y) = \int_{sky} (F_y^XF_y^Z) + (F_x^XF_x^Z) \\&= \int_{sky} (F_y^YF_y^Z) + (F_x^YF_x^Z)
      = -\frac{-3}{20} 
      \end{aligned}
        \end{equation}
    \item Scalar modes:
        \begin{equation}
        \begin{aligned}
                &\int_{sky} (F_b^X)^2 + (F_l^X)^2 = \int_{sky} (F_b^Y)^2 + (F_l^Y)^2 \\
                 &= \int_{sky} (F_b^Z)^2 + (F_l^Z)^2
      = \frac{1}{10} \\
               & \int_{sky} (F_l^XF_l^Y) + (F_b^XF_b^Y) = \int_{sky} (F_l^XF_l^Z) + (F_b^XF_b^Z) \\
      &= \int_{sky} (F_b^YF_b^Z) + (F_l^YF_l^Z)
      = -\frac{-1}{20} 
      \end{aligned}
        \end{equation}  
\end{itemize}

We can summarize the contribution from an isotropic SGWB to each channel for the different polarizations: 
\begin{itemize}
    
\item Vector modes:

\begin{equation}
\begin{aligned}
  \left<X(f)X^*(f')\right>&= \frac{1}{2}\delta(f-f')\left[ S_n^X(f) + \frac{3}{10}S_h(f)\right] \\
     \left<A(f)A^*(f')\right>&= \frac{1}{2}\delta(f-f')\left[ S_n^A(f) + \frac{9}{20}S_h(f)\right]  \\  
     \left<E(f)E^*(f')\right>&= \frac{1}{2}\delta(f-f')\left[ S_n^E(f) + \frac{9}{20}S_h(f)\right]  \\  
     \left<A(f)E^*(f')\right>&= \frac{1}{2}\delta(f-f')\left[ S_n^{AE}(f) + 0S_h(f)\right]  \\  
\end{aligned}
\end{equation}

 \item Scalar modes:

\begin{equation}
\begin{aligned}
  \left<X(f)X^*(f')\right>&= \frac{1}{2}\delta(f-f')\left[ S_n^X(f) + \frac{1}{10}S_h(f)\right] \\
     \left<A(f)A^*(f')\right>&= \frac{1}{2}\delta(f-f')\left[ S_n^A(f) + \frac{3}{20}S_h(f)\right]  \\  
     \left<E(f)E^*(f')\right>&= \frac{1}{2}\delta(f-f')\left[ S_n^E(f) + \frac{3}{20}S_h(f)\right]  \\  
     \left<A(f)E^*(f')\right>&= \frac{1}{2}\delta(f-f')\left[ S_n^{AE}(f) + 0S_h(f)\right]  \\  
\end{aligned}
\end{equation}

\end{itemize}

\end{document}